\documentclass[final]{cvpr}

\usepackage{times}
\usepackage{epsfig}
\usepackage{graphicx}
\usepackage{amsmath}
\usepackage{amssymb}
\usepackage{multirow}
\usepackage{soul}
\usepackage[export]{adjustbox}
\usepackage{xcolor,colortbl}
\definecolor{Gray}{gray}{0.85}

\usepackage{caption}
\usepackage{subcaption}
\usepackage{booktabs}
\usepackage{float}
\newcommand{\z}{\mathbf{z}}
\newcommand{\uh}{\mathbf{u}}
\newcommand{\y}{\mathbf{y}}
\newcommand{\x}{\mathbf{x}}
\newcommand{\n}{\mathbf{n}}

\newcommand{\xlr}{\x_{\text{LR}}}
\newcommand{\xhf}{\x_{\text{HF}}}
\newcommand{\zhf}{\mathbf{z}_{\text{HF}}}
\newcommand{\ngaussian}{\mathcal{N}(\mathbf{0}, \mathbf{I})}
\newcommand\norm[1]{\left\lVert#1\right\rVert}

\usepackage[colorlinks,linkcolor={blue},pagebackref=true,breaklinks=true,colorlinks,bookmarks=false]{hyperref}

\setstcolor{red}
\def\ourmodel{InvDN~}
\def\ourmodels{InvDN}

\def\cvprPaperID{603} % *** Enter the CVPR Paper ID here
\def\confYear{CVPR 2021}
\begin{document}

%%%%%%%%% TITLE
\title{Invertible Denoising Network: A Light Solution for Real Noise Removal}

\author{Yang Liu$^{1,2}$,  Zhenyue Qin$^{1}$\thanks{Corresponding authors: Zhenyue Qin (\textit{zhenyue.qin@anu.edu.au}) and Yang Liu (\textit{yang.liu3@anu.edu.au}).}, Saeed Anwar$^{1,2}$, Pan Ji$^{3}$, Dongwoo Kim$^{4}$, Sabrina Caldwell$^{1}$, Tom Gedeon$^{1}$\\
Australian National University$^1$, Data61-CSIRO$^2$, OPPO US Research$^3$, GSAI POSTECH$^4$\\
%{\tt\small \{yang.liu3, zhenyue.qin\}@anu.edu.au} \\
% \tt\small saeed.anwar@csiro.au, peterji530@gmail.com, dongwookim@postech.ac.kr}
}
% Yang Liu\\
% ANU
% \and
% \\
% ANU
% \and
% Saeed Anwar\\
% ANU/CSIRO
% \and
% Pan Ji\\
% OPPO US Research Center
% \and
% Dongwoo Kim\\
% POSTECH
% \and
% Sabrina Caldwell\\
% ANU
% \and
% Tom Gedeon\\
% ANU
% }

\maketitle

%%%%%%%%% ABSTRACT
\begin{abstract}
Invertible networks have various benefits for image denoising since they are lightweight, information-lossless, and memory-saving during back-propagation. However, applying invertible models to remove noise is challenging because the input is noisy, and the reversed output is clean, following two different distributions. We propose an invertible denoising network, \ourmodels, to address this challenge. \ourmodel transforms the noisy input into a low-resolution clean image and a latent representation containing noise. To discard noise and restore the clean image, \ourmodel replaces the noisy latent representation with another one sampled from a prior distribution during reversion. The denoising performance of \ourmodel is better than all the existing competitive models, achieving a new state-of-the-art result for the SIDD dataset while enjoying less run time. Moreover, the size of \ourmodel is far smaller, only having 4.2\% of the number of parameters compared to the most recently proposed DANet. Further, via manipulating the noisy latent representation, \ourmodel is also able to generate noise more similar to the original one. 
% reducing average KL divergence (AKLD) from 0.212 to 0.059.
Our code is available at: 
\href{https://github.com/Yang-Liu1082/InvDN.git}{https://github.com/Yang-Liu1082/InvDN.git}.
\end{abstract}

%%%%%%%%% BODY TEXT
\section{Introduction}
\begin{figure}[ht]
\centering
\begin{subfigure}{.15\textwidth}
  \centering
  % include first image
  \includegraphics[width=.99\linewidth]{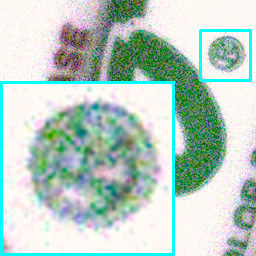}  
  \caption*{Noisy}
  \label{fig:sub-first}
\end{subfigure}
\begin{subfigure}{.15\textwidth}
  \centering
  % include second image
  \includegraphics[width=.99\linewidth]{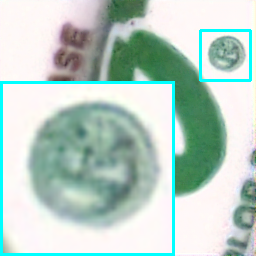}  
  \caption*{CBDNet~\cite{Guo2019Cbdnet}}
  \label{fig:sub-second}
\end{subfigure}
\begin{subfigure}{.15\textwidth}
  \centering
  % include second image
  \includegraphics[width=.99\linewidth]{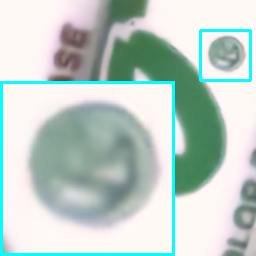}  
  \caption*{RIDNet~\cite{RIDNet}}
  \label{fig:sub-second}
\end{subfigure}
\begin{subfigure}{.15\textwidth}
  \centering
  % include first image
  \includegraphics[width=.99\linewidth]{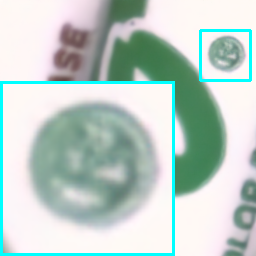}  
  \caption*{VDN~\cite{VDN}}
  \label{fig:sub-first}
\end{subfigure}
\begin{subfigure}{.15\textwidth}
  \centering
  % include second image
  \includegraphics[width=.99\linewidth]{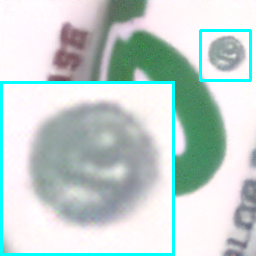}  
  \caption*{DANet~\cite{DANet}}
  \label{fig:sub-second}
\end{subfigure}
\begin{subfigure}{.15\textwidth}
  \centering
  % include second image
  \includegraphics[width=.99\linewidth]{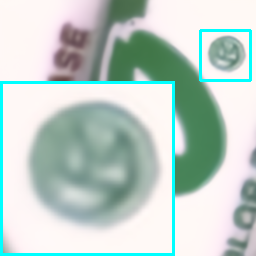}  
  \caption*{\ourmodel (Ours)}
  \label{fig:sub-second}
\end{subfigure}
% \vspace{-2mm}
\caption{A real noisy image from the SIDD~\cite{SIDD_2018_CVPR} dataset. Compared with RIDNet~\cite{RIDNet} and DANet~\cite{DANet}, \ourmodel does not over-smooth. In addition, in comparison with all the other methods, \ourmodel restores more crisp edges and produces fewer artifacts. The examples are best viewed in color on a high-resolution display with zooming in. }
% \vspace{-6mm}
\label{fig:RNI_Front}
\end{figure}

% Image Denoising is an important task
Image denoising aims to restore clean images from noisy observations. Traditional approaches model denoising as a maximum a posteriori (MAP) optimization problem, with assumptions on the distribution of noise~\cite{6751276, 7557070, 7410532}, and natural image priors~\cite{4271520, 6122031,4060941}. Although these algorithms achieve satisfactory performance on removing synthetic noise, their effectiveness on real-world noise is compromised since their assumptions deviate from those in real-world scenarios. 
% do not generally hold for real-world scenarios. 
Recently, convolutional neural networks (CNNs) have achieved superior denoising performance~\cite{DnCNN, zhang2018ffdnet}. These CNNs learn the features of images from a large number of clean and noisy image pairs. However, since real noise is very complex, to achieve better denoising accuracy, CNN denoising models have become increasingly large and complicated~\cite{VDN, DANet, Zamir2020MIRNet}. % to reconstruct the clean ground-truth.
Thus, although some methods can achieve very impressive denoising results, they may not be practical in realistic scenarios such as deploying the model on  edge equipment like smartphones and motion sensing devices.

Currently, a substantial amount of research has been devoted to developing neural networks that are invertible~\cite{dinh2014nice,rezende2015variational_flow,jacobsen2018irevnet,pmlr-v97-behrmann19a}. For image denoising, invertible networks are advantageous from the following three aspects: (1) the model is light, as encoding and decoding use the same parameters; (2) they preserve details of the input data since invertible networks are information-lossless~\cite{liu2020deep}; (3) they save memory during back-propagation because they use a constant amount of memory to compute gradients, regardless of the depth of the network~\cite{gomez2017reversible}. Hence, invertible models are suitable for small devices like smartphones. We thus study employing invertible networks to address the problem of image denoising. However, applying such networks to remove noise is non-trivial. The original inputs and the reversed results of the traditional invertible models follow the same distribution~\cite{dinh2017RealNVP,kingma2018glow, xiao2020invertible}. In contrast, for image denoising, the input is noisy, and the restored image is clean, following two different distributions. Therefore, invertible denoising networks are required to abandon the noise in the latent space before the reversion. Due to this difficulty, noise removal has not previously been studied and deployed in invertible literature and models. 

In this paper, we propose an invertible denoising network, InvDN, to resolve the above difficulties. Unlike previous invertible models, two different latent variables are involved; one incorporates noise and high-frequency clean contents while the other only encodes the clean part. During the forward pass, \ourmodel transforms the input image to a downscaled latent representation with an increased number of channels. We train \ourmodel to make the first three channels of the latent representation the same as the low-resolution clean image. Since invertible networks preserve all the information of the input~\cite{liu2020deep}, noisy signals are in the rest of the channels. To remove noise completely, we discard all the channels that contain noise. However, as a side-effect, we also lose some information corresponding to the high-resolution clean image. To reconstruct such missing information, we sample a new latent variable from a prior distribution and combine it with the low-resolution image to restore the clean image. 

Our contributions are as follows: 
\begin{itemize}
    \setlength\itemsep{1mm}
    \item We are the first to design invertible networks for real image denoising to the best of our knowledge.
    \item The latent variable of traditional invertible networks follows a single distribution. Instead, \ourmodel has two latent variables following two different distributions. Thus, \ourmodel can not only restore clean images but also generate new noisy images. 
    \item We achieve a new state-of-the-art (SOTA) result on the SIDD test set, using far fewer parameters and less run time than the previous SOTA methods.
    \item \ourmodel is able to generate new noisy images that are more similar to the original noisy ones. 
\end{itemize}

\section{Related Work}
In this section, we summarize and discuss the development and recent trends in image denoising. The widely used denoising methods can be classified into traditional methods and current data-driven deep learning methods.

\textbf{Traditional Methods.}
Model-driven denoising methods usually construct a MAP optimization problem with a loss and a regularization term. 
The assumptions on the noise distribution are needed for most traditional methods to build the model. One assumed distribution is the Mixture of Gaussian, which is used as an approximator for noise on natural patches or patch groups~\cite{6126278, 7410433, 7410393}.
The regularization term is usually based on the clean image's prior. Total variation~\cite{1992TotalVariation} denoising uses the statistical characteristics of images to remove noise. Sparsity~\cite{5459452} is enforced in dictionary learning methods~\cite{5995478}, to learn over-complete dictionaries from clean images. Non-local similarity~\cite{1467423, 4154791} methods employ non-local patches that share similar patterns.
% with a specific patch. 
Such a strategy is adopted by the notable BM3D~\cite{BM3D} and NLM~\cite{1467423}. 
However, these models are limited due to the assumptions on the prior of spatially invariant noise or clean images, which are often different from real cases, where the noise is spatially variant.

\textbf{Data-driven Deep Learning Denoising.}
Recent years have seen rapid progress in the deep learning methods, boosting the denoising performance to a large extent. The early deep models focus on synthetic noisy image denoising due to a lack of real data. As some large real noise datasets, such as DND~\cite{DND_2017_CVPR} and SIDD~\cite{SIDD_2018_CVPR}, have been presented, current research focuses on blind real image denoising. 
There are two main streams in real image denoising. One is to adapt the methods that work well on the synthetic datasets to the real datasets while considering the gap between these two domains~\cite{zhang2018ffdnet, Guo2019Cbdnet}. The current most competitive method along this direction is AINDNet~\cite{AINDNet}, which applies transfer learning from synthetic to real denoising with the Adaptive Instance Normalization operations.

The other direction is to model real noise with more complicated distributions and design new network architectures~\cite{ULRD,GCBD}. VDN~\cite{VDN}
proposed by Yue~\etal assumes that noise follows an inverse Gamma distribution, and the clean image we observe is a conjugate Gaussian prior of the unavailable real clean images. They propose a new training objective based on these assumptions and use two parallel branches to learn these two distributions in the same network. 
Its potential limitation is that the assumptions are not suitable when the noise distribution becomes complicated.
Later, DANet~\cite{DANet} abandons the assumptions for noise distributions and employs a GAN framework to train the model. Two parallel branches are also employed in this architecture: one for denoising and the other for noise generation. This design concept is that the three kinds of image pairs (clean and noisy, clean and generated noisy, as well as denoised and noisy) follow the same distribution, so they use a discriminator to train the model. 
The potential limitation is that GAN-based models' training is unstable and thus takes longer to converge~\cite{arjovsky2017wasserstein}. 
Furthermore, both VDN and DANet employ Unet~\cite{Unet_2015} in the parallel branches, making their models very large.

To compress the model size, we explore invertible networks. To the best of our knowledge, few studies apply invertible networks in denoising literature. Noise Flow~\cite{9008378} introduces an invertible architecture to learn real noise distributions to generate real noise as a way of data augmentation. Generating noisy images with Noise Flow requires extra information apart from the sRGB images, including raw-RGB images, ISO, and camera-specific values. They do not propose new denoising backbones. So far, no invertible network for real image denoising has been reported. 

\section{Invertible Denoising Network}
In this paper, we present a novel denoising architecture consisting of invertible modules, \ie, Invertible Denoising Network (InvDN). For completeness, in this section, we first provide the background of invertible neural networks and then present the details of InvDN.

\subsection{Invertible Neural Network}
Invertible networks are originally designed for unsupervised learning of probabilistic models~\cite{dinh2017RealNVP}. These networks can transform a distribution to another distribution through a bijective function without losing information~\cite{liu2020deep}. Thus, it can learn the exact density of observations. Using invertible networks, images following a complex distribution can be generated through mapping a given latent variable $\z$, which follows a simple distribution $p_{\z}(\z)$, to an image instance $\x \sim p_{\x}(\x)$, \ie, $\x = f(\z)$, where $f$ is the bijective function learned by the network. 
Due to the bijective mapping and exact density estimation properties,
% Due to their accurate inference ability, 
invertible networks have received increasing attention in recent years and have been applied successfully in applications such as image generation~\cite{dinh2017RealNVP, kingma2018glow} and rescaling~\cite{xiao2020invertible}.

% The invertible module we follow in this work is the coupling layer~\cite{dinh2017RealNVP}. Suppose the module's input is $\z^i$, and output is $\z^{i+1}$. This module's operations in the forward and backward pass are listed in \autoref{Tab:Invertible_Module}. 
% In R2 and R3, during the backward pass, only the \textit{Plus} ($+$) and \textit{Multiply} ($\odot$) operations are inverted to \textit{Minus} ($-$) and \textit{Divide} ($/$) while the operations performed by $\operatorname{\phi}_1$, $\operatorname{\phi}_2$, $\operatorname{\phi}_3$ and $\operatorname{\phi}_4$
% are not required to be invertible.
% $\operatorname{\phi}_1$, $\operatorname{\phi}_2$, $\operatorname{\phi}_3$ and $\operatorname{\phi}_4$ can be any networks.

\subsection{Challenges in Denoising with Invertible Models}
\label{subsec:concep_of_design}
Applying invertible models in denoising is different from other applications. The widely used invertible networks employed in image generation~\cite{dinh2017RealNVP, kingma2018glow} and rescaling~\cite{xiao2020invertible} consider the input and the reverted image to follow the same distribution. Such applications are straightforward candidates for invertible models. Image denoising, however, takes a noisy image as input and reconstructs a clean one, \ie{}, the input and the reverted outcome follow two different distributions. 
% Thus, it is non-trivial to implement denoising with invertible architectures.
On the other hand, an invertible transform does not lose any information during the transformation. However, the lossless property of invertibility is not desired for image denoising since the noise information remains while we transform an input image into latent variables. If we can disentangle the noisy and clean signals during an invertible transformation, we may reconstruct a clean image without worrying about losing any important information by abandoning the noisy information. In the following section, we present one way to obtain a clean signal through an invertible transformation.

\subsection{Concept of Design}
We denote the original noisy image as $\y$, its clean version as $\x$ and the noise as $\n$. We have: $p(\y) = p(\x, \n) = p(\x) p(\n | \x)$. Using the invertible network, the learned latent representation of observation $\y$ contains both noise and clean information. It should be noted that it is non-trivial to disentangle them and abandon only the noisy part.

\begin{figure}
\centering
\includegraphics[width=0.48\textwidth]{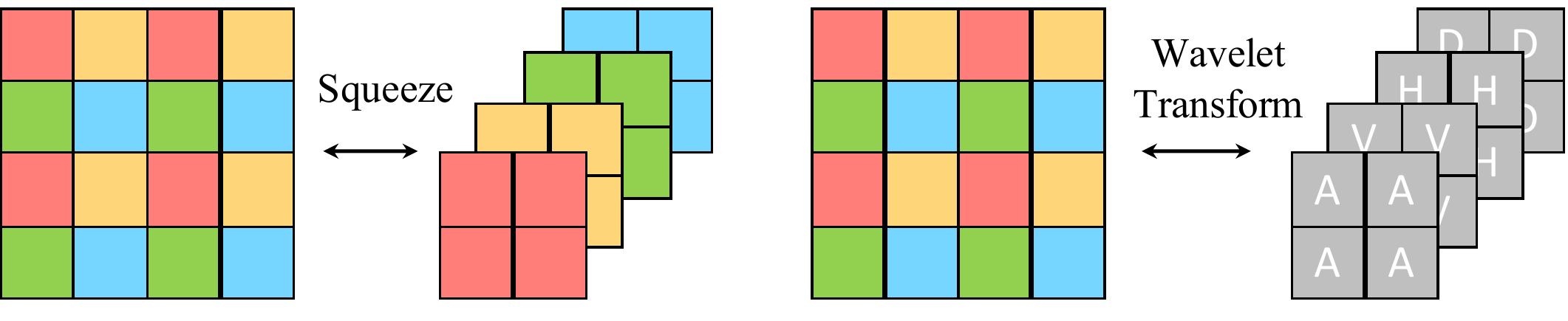}
\caption{Invertible Transforms. The Squeeze operation on the left extracts features according to a checkerboard pattern. The Wavelet Transformation on the right extracts the average pooling of the image, as well as the vertical, horizontal and diagonal derivatives.}   
\label{fig:Transform}
% \vspace{-3mm}
\end{figure}

On the other hand, invertible networks utilize different feature extraction approaches comparing with existing deep denoising models.
Existing ones usually employ convolutional layers with padding to extract features. However, they are not invertible due to two reasons: Firstly, the padding makes the network non-invertible; secondly, the parameter matrices of convolutions may not be full-rank. 
Thus, to ensure invertibility, rather than using convolutional layers, it is necessary to utilize invertible feature extraction methods, such as the Squeeze layer~\cite{kingma2018glow} and Haar Wavelet Transformation~\cite{ardizzone2019guided}, as presented in \autoref{fig:Transform}. The Squeeze operation reshapes the input into feature maps with more channels according to a checkerboard pattern. Haar Wavelet Transformation extracts the average pooling of the original input as well as the vertical, horizontal, and diagonal derivatives. As a result, the spatial size of the feature maps extracted by the invertible methods is inevitably downscaled.
% , with an increase in the number of channels.  

\begin{figure*}
\centering
\includegraphics[width=\textwidth]{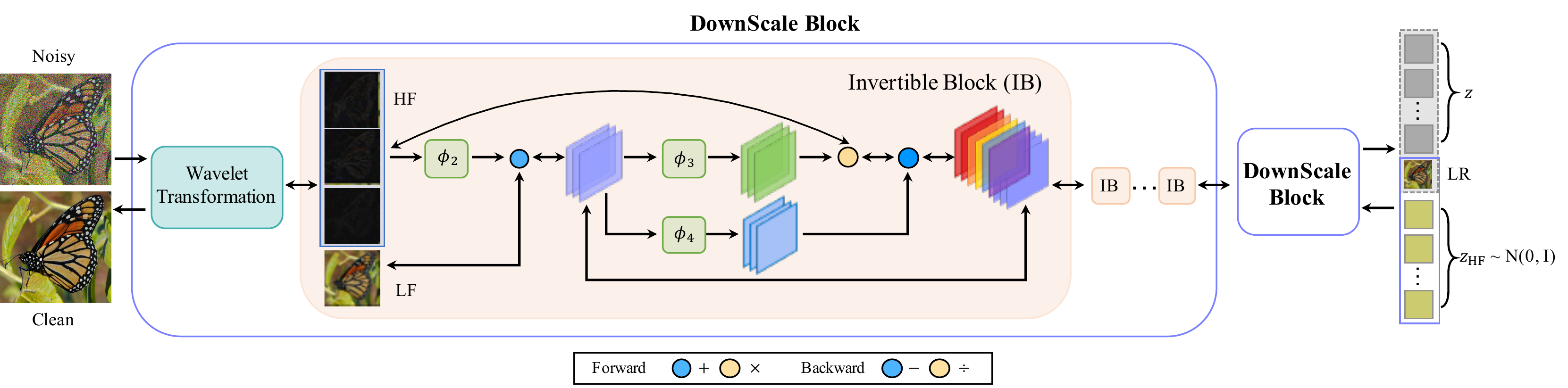}
\caption{Invertible Denoising Network. \ourmodel consists of several DownScale Blocks. Each DownScale Block has a Wavelet Transformation layer to reduce the spatial resolution by 2$\times$ and increase the number of channels by 4$\times$, followed by several Invertible Blocks. In the forward pass, we learn the low-resolution image and a latent variable $\z$ from the noisy input. In the backward procedure, we randomly sample $\zhf$ from $\ngaussian$ and combine it with LR to restore the clean image.}   
\label{fig:network}
% \vspace{-2mm}
\end{figure*}

Therefore, instead of disentangling the clean and noisy signals directly, we aim to separate the low-resolution and high-frequency components of a noisy image. 
The sampling theory~\cite{gastal2017spectral} indicates that during the downsampling process, the high-frequency signals are discarded. Since invertible networks are information-lossless~\cite{liu2020deep}, if we make the first three channels of the transformed latent representation to be the same as the downsampled clean image, high-frequency information will be encoded in the remaining channels. 
% If we use the low-resolution clean image to supervise the invertible transformation from the input image to part of its latent representation, this transformation can be treated as a downsampling process. 
% According to the sampling theory~\cite{gastal2017spectral}, high-frequency information will be encoded in the remaining part of the latent representation during downsampling. 
Based on the observation that the high-frequency information contains noise as well, we abandon all high-frequency representations 
% from the transformation 
before inversion
to reconstruct a clean image from low-resolution components. We formally describe the process as follows: 
\begin{equation*}
    p(\y) = p(\xlr, \xhf, \n) = p(\xlr) p(\xhf, \n | \xlr),
\end{equation*}
where $\xlr$ represents the low-resolution clean image.
% We train the first three channels of the downscaled representation to be the same as the low-resolution clean image. Then, the remaining channels encode residual information, including the noise. 
% According to the sampling theory~\cite{gastal2017spectral}, high-frequency information will lose during downsampling. 
We use $\xhf$ to represent the high-frequency contents that cannot be obtained by $\xlr$ when reconstructing the original clean image. Since it is challenging to disentangle $\xhf$ and $\n$, we abandon all the channels representing $\z \sim p(\xhf, \n | \xlr)$ to remove spatially variant noise completely. Nevertheless, a side-effect is the loss of $\xhf$. 
To reconstruct $\xhf$, we sample $\zhf \sim \ngaussian$ and train our invertible network to transform $\zhf$, in conjunction with $\xlr$, to restore the clean image $\x$. In this way, the lost high-frequency clean details $\xhf$ is embedded in the latent variable $\zhf$.

\subsection{Network Architecture}
\label{subsec:network}
We first employ a supervised approach to guide the network to separate high-frequency and low-resolution components during transformation. 
% Let $\x$ be an input image and $\xhf$ and $\xlr$ be high-frequency and low-frequency components obtained by some invertible transformation $f$, \ie $f(\x) = [\xlr;\xhf]$.
After some invertible transformation $g$, the noisy image $\y$ is transformed into its corresponding low-resolution clean image and high-frequency encoding $\z$, \ie $g(\y) = [g(\y)_{\text{LR}};\z]$.
% We denote $f(\x)_{\text{LR}}$ to represent low-frequency components $\xlr$ and $f(\x)_{\text{HF}}$ high frequency components $\xhf$ of $\x$. 
% To make the network separate $\xhf$ and $\xlr$ from $\x$, we minimize the following forward objective
% To separate $\xlr$ and $\z$, 
We minimize the following forward objective
\begin{align}\label{loss:forw}
    \mathcal{L}_{\text{forw}}(\y, \mathbf{x}_{\text{LR}}) &= \frac{1}{M}\sum_{i=1}^{M} \norm{ g(\y)_{\text{LR}} - \mathbf{x}_{\text{LR}} }_m,
\end{align}
where $g(\y)_{\text{LR}}$ is the low-frequency components learned by the network, corresponding to three channels of the output representation in the forward pass. $M$ is the number of pixels. $||\cdot||_m$ is the $m$-norm and $m$ can be either 1 or 2. To obtain the ground truth low-resolution image $\xlr$, we down-sample the clean image $\x$ via bicubic transformation.

% To denoise the original image with the low-resolution image, 
To restore the clean image with $g(\y)_{\text{LR}}$, 
we use inverse transform $g^{-1}([g(\y)_{\text{LR}};\zhf])$ with random variable $\zhf$ sampled from normal distribution $\mathcal{N}(\mathbf{0},\mathbf{I})$.
% we use $f(\y)_{\text{LR}}$ via inverse transform $f^{-1}([f(\y)_{\text{LR}};\zhf])$ with random variable $\zhf$ following the standard normal distribution $\mathcal{N}(\mathbf{0},\mathbf{I})$. 
The backward objective is written as
\begin{align*}\label{loss:back}
    \mathcal{L}_{\text{back}}(g(\y)_{\text{LR}}, \mathbf{x}) &= \frac{1}{N}\sum_{i=1}^{N} \norm{ g^{-1}([g(\y)_{\text{LR}};\zhf]) - \mathbf{x} }_m, 
\end{align*}
where $\x$ is the clean image.
% and $\zhf \sim \mathcal{N}(\mathbf{0},\mathbf{I})$. 
$N$ is the number of pixels.
% During the training, we randomly sample $\z$ from the standard normal. Hence, to reduce the backward objective, the invertible transform $f$ requires to make $\xhf$ follow the standard normal distribution.
We train the invertible transformation $g$ by simultaneously utilizing both forward and backward objectives.

Inspired by~\cite{xiao2020invertible, ardizzone2019guided, dinh2017RealNVP}, the invertible transform $g$ we present is of a multi-scale architecture, consisting of several down-scale blocks. Each down-scale block consists of an invertible wavelet transformation followed by a series of invertible blocks. The overall architecture of the \ourmodels{} model is demonstrated in \autoref{fig:network}.

\begin{table*}[t!]
\centering
\caption{The invertible block operations in the down-scale block. 
In the forward pass, the input and output to each block are denoted as $\uh^i$ and $\uh^{i+1}$. 
% Suppose the block's input is $\uh^i$, and output is $\uh^{i+1}$. 
During the backward pass, only \textit{Plus} ($+$) and \textit{Multiply} ($\odot$) needs to be inverted. $\operatorname{\phi}_1$, $\operatorname{\phi}_2$, $\operatorname{\phi}_3$ and $\operatorname{\phi}_4$ do not need to be inverted. Thus, they can be any neural networks. }
% \vspace{-2mm}
\renewcommand{\arraystretch}{1.6}
\resizebox{\textwidth}{!}{
\begin{tabular}{l|c|c|l}
\hline
\hline
\# & Forward Operation & Backward Operation & Specification \\ 
\hline
R1 & \multirow{1}{*}{$\uh^i_{a}$, $\uh^i_{b} = \operatorname{Split}(\uh^i)$} & \multirow{1}{*}{$\uh^{i+1}_{a}$, $\uh^{i+1}_{b} = \operatorname{Split}(\uh^{i+1})$} & \multirow{1}{18em}{$\operatorname{Split}(\cdot)$ is splitting channel-wise.}\\ \hline 
R2 & $\uh^{i+1}_{a} = \uh^i_{a} \odot \operatorname{exp}(\operatorname{\phi}_1(\uh^i_{b})) + \operatorname{\phi}_2(\uh^i_{b})$ & $ \uh^i_{b} = (\uh^{i+1}_{b} - \operatorname{\phi}_4(\uh^{i+1}_{a})) / \operatorname{exp}(\operatorname{\phi}_3(\uh^{i+1}_{a})$) & \multirow{2}{19em}{$\operatorname{\phi}_1$, $\operatorname{\phi}_2$, $\operatorname{\phi}_3$ and $\operatorname{\phi}_4$ can be any neural networks. \\ $\odot$ is the multiply operation.}  \\ 
\cline{1-3}
% 3 & $u^{i+1}_b = u^i_{b}$ & $ u^i_{b} = u^{i+1}_{b}$ &  \\ \hline
R3 & $\uh^{i+1}_b = \uh^i_{b} \odot \operatorname{exp}(\operatorname{\phi}_3(\uh^{i+1}_{a})) + \operatorname{\phi}_4(\uh^{i+1}_{a})$ & $ \uh^i_{a} = (\uh^{i+1}_{a} - \operatorname{\phi}_2(\uh^{i}_{b})) /
\operatorname{exp}(\operatorname{\phi}_1(\uh^i_{b})$) &  \\ \hline
R4 & \multirow{1}{*}{$\uh^{i+1} = \operatorname{Concat}(\uh^{i+1}_{a}, \uh^{i+1}_{b})$} & \multirow{1}{*}{$\uh^i = \operatorname{Concat}(\uh^i_{a}$, $\uh^i_{b})$} & \multirow{1}{18em}{$\operatorname{Concat}(\cdot)$ is concating channel-wise.}\\ \hline
% & &  \\ \hline
\hline
\end{tabular}
}
% \vspace{-2mm}
\label{Tab:Invertible_Module}
\end{table*}
\renewcommand{\arraystretch}{1}

\textbf{Invertible Wavelet Transformation.}
% \textbf{The \st{wavelet} transformation.}
Since we aim to learn the low-resolution clean image in the forward pass, we apply invertible discrete wavelet transformations (DWTs), specifically Haar wavelet transformation, to increase feature channels and to down-sample feature maps. After the wavelet transformation, the input image or an intermediate feature map with the size of $(H, W, C)$ is transformed into a new feature map of size $(H/2, W/2, 4C)$.
Haar wavelets decompose an input image into one low-frequency representation and three high-frequency representations in the vertical, horizontal, and diagonal direction~\cite{Kingsbury1998}. Other DWTs can also be exploited, such as Haar, Daubechies, and Coiflet wavelets~\cite{xiao2020invertible,ardizzone2019guided}.

\textbf{Invertible Block.}
The wavelet transformation layer in each down-scale block splits the representation into low- and high-frequency signals, which are further processed by a series of invertible blocks. 
% An invertible block consists of four sequences of invertible operations presented in \autoref{Tab:Invertible_Module}.
%Next, we describe the implementation of each operation in our network.  
The invertible block we follow in this work is the coupling layer~\cite{dinh2017RealNVP}. Suppose the block's input is $\uh^i$, and output is $\uh^{i+1}$. This block's operations in the forward and backward pass are listed in~\autoref{Tab:Invertible_Module}. 

$\operatorname{Split}(\cdot)$ operation divides input feature map $\uh^i$, which has the size of $(H/2, W/2, 4C)$, into 
$\uh_a^i$ and $\uh_b^i$, corresponding to the low-frequency image representations of size $(H/2, W/2, C)$ and the high-frequency features (such as texture and noise) of size $(H/2, W/2, 3C)$, respectively.

In R2 and R3 of~\autoref{Tab:Invertible_Module}, during the backward pass, only the \textit{Plus} ($+$) and \textit{Multiply} ($\odot$) operations are inverted to \textit{Minus} ($-$) and \textit{Divide} ($/$) while the operations performed by $\operatorname{\phi}_1$, $\operatorname{\phi}_2$, $\operatorname{\phi}_3$ and $\operatorname{\phi}_4$
are not required to be invertible.
Thus, $\operatorname{\phi}_1$, $\operatorname{\phi}_2$, $\operatorname{\phi}_3$ and $\operatorname{\phi}_4$ can be any networks, including convolutional layers with paddings.
Since the skip connection is shown to be crucial for deep denoising networks~\cite{VDN, DANet}, we simplify the forward operation in R2 as
% choose the identity function as $\operatorname{\phi}_1$. The operation in R2 is simplified as
\begin{equation}
  \mathbf{u}^{i+1}_{a} = \mathbf{u}^i_{a} + \operatorname{\phi}_2(\mathbf{u}^i_{b}).
\end{equation}
The low-frequency features can be passed to deep layers with this approach. We use the residual block as operations $\operatorname{\phi}_2$, $\operatorname{\phi}_3$ and $\operatorname{\phi}_4$. 

$\operatorname{Concat}(\cdot)$ is the inverse operation of $\operatorname{Split}(\cdot)$, which concatenates the feature maps along channels and passes them into the next module.

\section{Experiment}
\begin{figure*}[t]
\centering
% Caption
\begin{subfigure}{.16\textwidth}
  \centering
  % include first image
  \caption*{Noisy} 
\end{subfigure}
\begin{subfigure}{.16\textwidth}
  \centering
  % include second image
  \caption*{CBDNet~\cite{Guo2019Cbdnet}} 
\end{subfigure}
\begin{subfigure}{.16\textwidth}
  \centering
  % include second image
  \caption*{RIDNet~\cite{RIDNet}} 
\end{subfigure}
\begin{subfigure}{.16\textwidth}
  \centering
  % include first image
  \caption*{VDN~\cite{VDN}}   
\end{subfigure}
\begin{subfigure}{.16\textwidth}
  \centering
  % include second image
  \caption*{DANet~\cite{DANet}} 
\end{subfigure}
\begin{subfigure}{.16\textwidth}
  \centering
  % include second image
  \caption*{\ourmodel (Ours)} 
\end{subfigure}
% SIDD
\begin{subfigure}{1.\textwidth} 
  \centering
  \includegraphics[width=.16\linewidth]{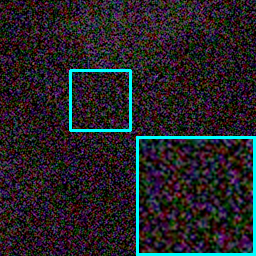} 
  \includegraphics[width=.16\linewidth]{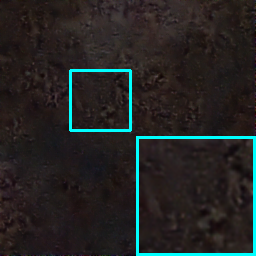}  
  \includegraphics[width=.16\linewidth]{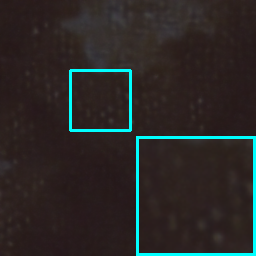} 
  \includegraphics[width=.16\linewidth]{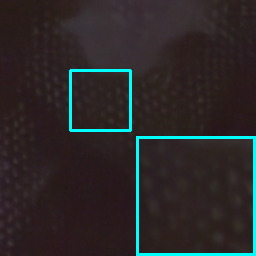}  
  \includegraphics[width=.16\linewidth]{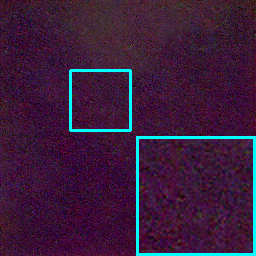}  
  \includegraphics[width=.16\linewidth]{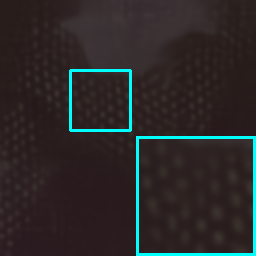}
  \caption{\textbf{SIDD~\cite{SIDD_2018_CVPR} visualization examples.} \ourmodel produces fewer artifacts and more clear patterns, even in dark environments. }
  \label{fig:sub-first}
\end{subfigure}

% DnD
\begin{subfigure}{1.\textwidth} 
  \centering
  \includegraphics[width=.16\linewidth]{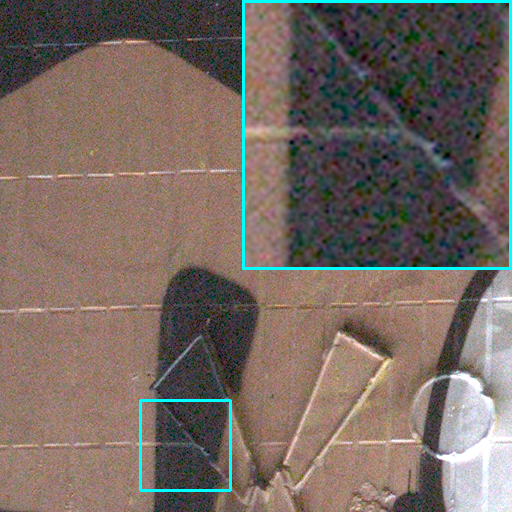} 
  \includegraphics[width=.16\linewidth]{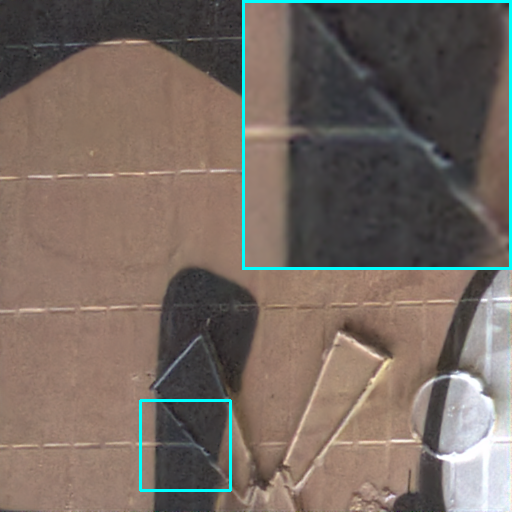}  
  \includegraphics[width=.16\linewidth]{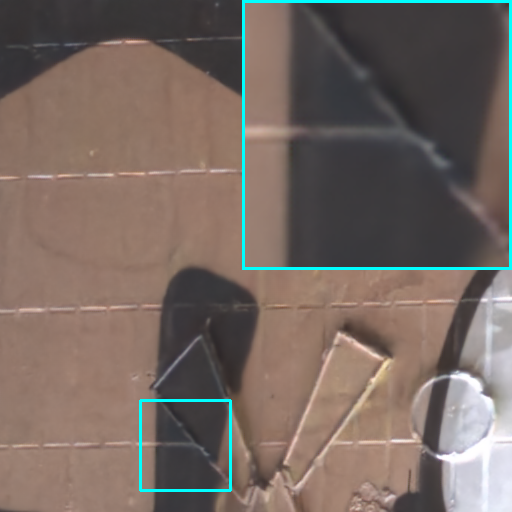} 
  \includegraphics[width=.16\linewidth]{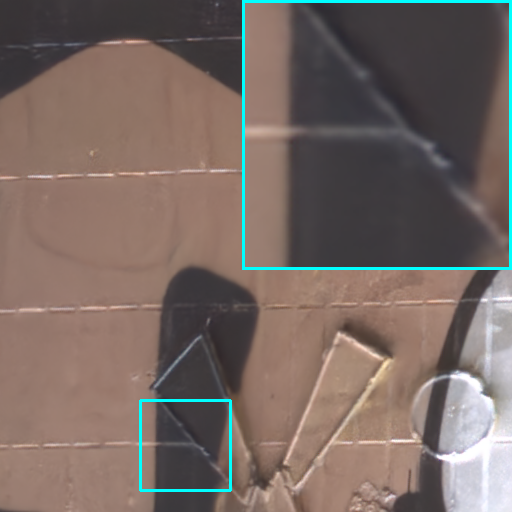}  
  \includegraphics[width=.16\linewidth]{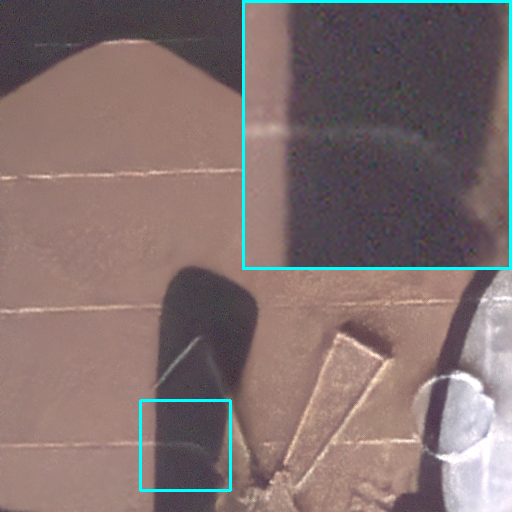}  
  \includegraphics[width=.16\linewidth]{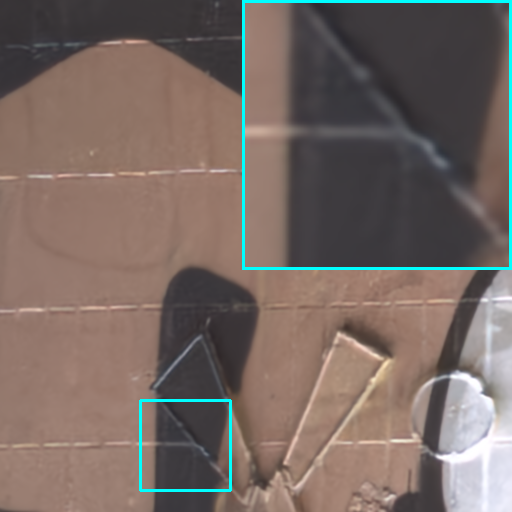}
  \caption{\textbf{DND~\cite{DND_2017_CVPR} visualization examples.} \ourmodel can reconstruct the subtle edges very clearly}
%   produces fewer fake dots, restores better colors and finer textures. }
  \label{fig:sub-first}
\end{subfigure}

% RNI
\begin{subfigure}{1.\textwidth} 
  \centering
  \includegraphics[width=.16\linewidth]{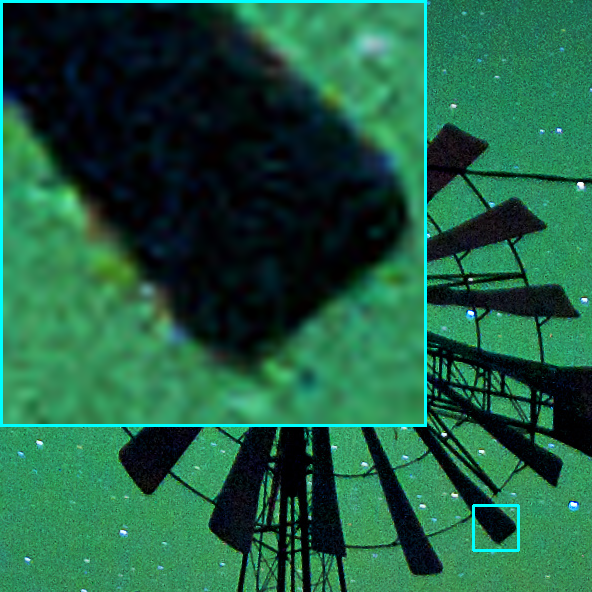} 
  \includegraphics[width=.16\linewidth]{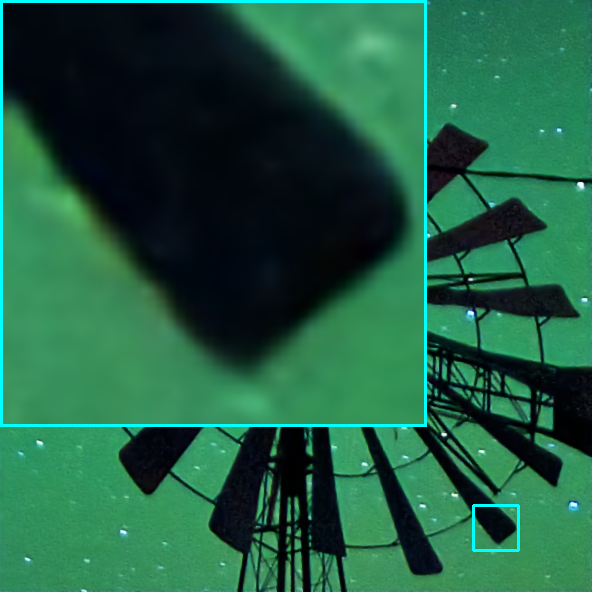}  
  \includegraphics[width=.16\linewidth]{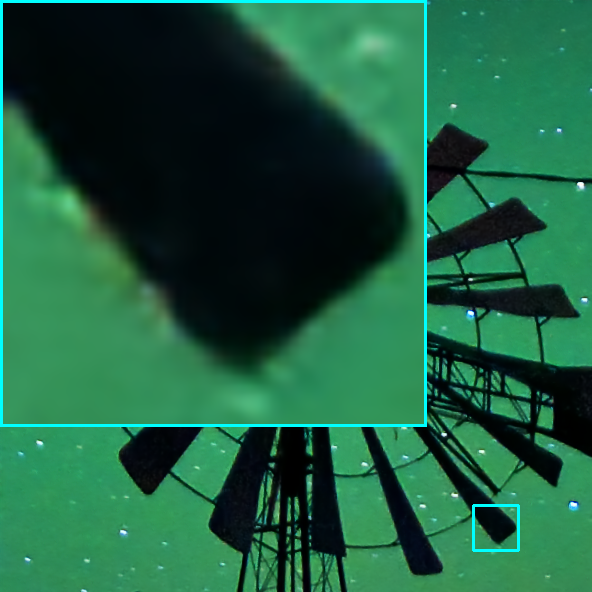} 
  \includegraphics[width=.16\linewidth]{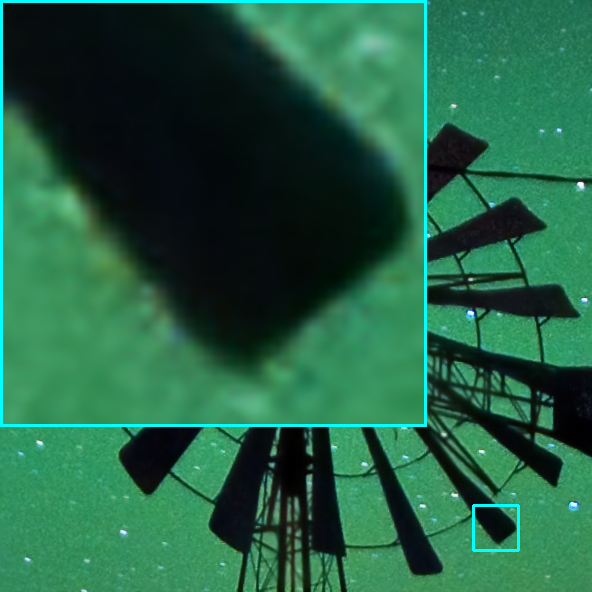}  
  \includegraphics[width=.16\linewidth]{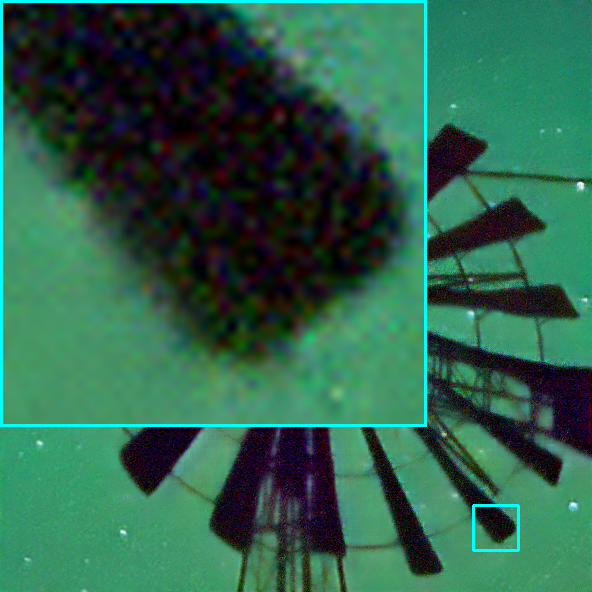}  
  \includegraphics[width=.16\linewidth]{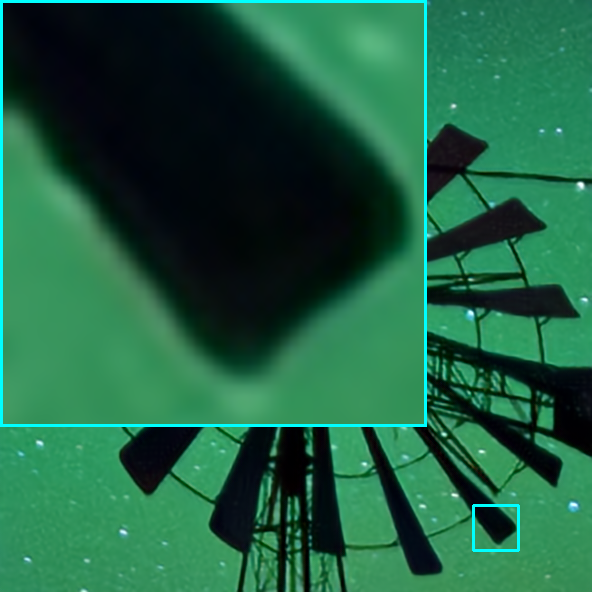}
  \caption{\textbf{RNI visualization examples.} \ourmodel produces fewer stains and more accurate edges without artifacts.}
%   , and more accurate textures. }
  \label{fig:sub-first}
\end{subfigure}
% \vspace{-2mm}
\caption{Visualization results of \ourmodel compared against other competitive models. Boxed regions are zoomed results. Best viewed in color on a high-resolution display. }
\label{fig:visual_rsts}
% \vspace{-6mm}
\end{figure*}

In this section, we present some empirical performances of \ourmodels{} on real denoising tasks.

\subsection{Dataset}
We evaluate \ourmodel on three real-world denoising benchmarks, \ie, SIDD~\cite{SIDD_2018_CVPR}, DND~\cite{DND_2017_CVPR}, and RNI~\cite{RNI15}. 

%\noindent 
\textbf{SIDD}~\cite{SIDD_2018_CVPR} is taken by five smartphone cameras with small apertures and sensor sizes. We use the medium version of SIDD as the training set, containing 320 clean-noisy pairs for training and 1280 cropped patches from the other 40 pairs for validation. In each iteration of training, we crop the input image into multiple $144 \times 144$ patches to feed into the network. The reported test results are obtained via an online submission system. 

%\noindent 
\textbf{DND}~\cite{DND_2017_CVPR} is captured by four consumer-grade cameras of differing sensor sizes. It contains 50 pairs of real-world noisy and approximately noise-free images. These images are cropped into 1000 patches of size $512\times512$. Similarly to SIDD, the performance is evaluated by submitting the outputs of the methods to the online system. 

%\noindent 
\textbf{RNI15}~\cite{RNI15} is composed of 15 real-world noisy images without ground-truths. Therefore, we only provide visual comparisons on this dataset.

\subsection{Training Details.}
Our \ourmodel has two down-scale blocks, each of which is composed of eight invertible blocks. All the models are trained with Adam~\cite{kingma2015adam} as the optimizer, with momentum of $\beta_1=0.9, \beta_2=0.999$. The batch size is set as 14, and the initial learning rate is fixed at $2\times10^{-4}$, which decays by half every 50k iterations. PyTorch is used as the implementation framework, and training is performed on a single 2080-Ti GPU.  We augment the data with horizontal and vertical flipping, as well as random rotations of $90\times\theta$ where $\theta=0,1,2,3$.

\subsection{Experimental Results}
We evaluate the methods by commonly used metrics such as Peak Signal-to-Noise Ratio (PSNR) and Structural Similarity Index Measure (SSIM), which are also available on the real-noisy DND and SIDD websites.

\textbf{Quantitative Measure.}
As mentioned before, we train \ourmodel using the SIDD medium training set.
DND does not provide the training set; therefore, we use the model trained on SIDD. 
For a fair comparison, the PSNR of other competitive models on the test set is directly taken from the official DND and SIDD leaderboards and verified from the respective articles. \autoref{Tab:DnD_SIDD} reports the test results of different denoising models. We can observe that the number of model parameters is directly positively correlated to the model's performance. From RIDNet~\cite{RIDNet} to DANet~\cite{DANet}, the number of model parameters increases from 1.49 million (M) to 63.01M with a slight improvement in PSNR. 

Nevertheless, \ourmodel reverses the trend, having only 2.64M parameters. Compared with the most recently proposed SOTA DANet, \ourmodel only uses less than  4.2\% of the number of parameters of DANet. Although \ourmodel has far fewer parameters, the denoising performance on the test set is better than all the recently proposed denoising models, achieving a new SOTA result for the SIDD dataset\footnote{We only compare with the published methods trained on the benchmark SIDD training set.}. The performance of \ourmodel is also comparable with that of the recent competitive models on DND, indicating the generalization ability of our lightweight denoising model. 

\textbf{Qualitative Measure.}
To further illustrate the better performance of \ourmodel against other methods, in \autoref{fig:visual_rsts}, we show denoised visual results on images from three different datasets, SIDD, DND and RNI15.
The first row of the figure illustrates that \ourmodel restores well-shaped patterns, while other competing techniques induce blurry textures and artifacts. Furthermore, the second row depicts that \ourmodel recovers the subtle edges very clearly, whereas other models bring artifacts and over-smoothness. Finally, from the third row, we observe that \ourmodel reconstructs accurate edges compared to other models that introduce blockiness, fuzziness, and random dots, particularly along the edges. 
% The first row of the figure illustrates that \ourmodel restores more crisp edges, while other competing techniques induce zigzag or blurry edges. Furthermore, the second row depicts that \ourmodel recovers the actual structures and produces well-shaped edges, whereas other models induce artifacts. Finally, from the fifth and sixth rows, we observe that \ourmodel reconstructs accurate textures compared to other models that introduce blockiness, fuzziness, and random dots, particularly along the edges. 

\begin{table*}[t]
\caption{Comparison of \ourmodels{} against other competitive algorithms. All the results are obtained without MC-ensemble for fair comparison. ``Infer time'' presents the inference time of one $256 \times 256$ image in gigaFLOPs. A tick represents containing the corresponding functionality. The best results are emphasized with red. }
% \vspace{-2mm}
\centering
\resizebox{\linewidth}{!}{
% \begin{centering}
\begin{tabular}{|l|c|c|c|c|c|c|cc|cc|cc|}
\hline
\rowcolor{Gray}
 &  &  & &  &  & & \multicolumn{2}{c|}{DND} & \multicolumn{2}{c|}{SIDD}\\
\cline{8-11}
\rowcolor{Gray}
\multirow{-2}{2em}{Method} & \multirow{-2}{4.5em}{Blind Denoising} & \multirow{-2}{*}{MultiScale} & \multirow{-2}{4.5em}{Noise Generation} & \multirow{-2}{*}{Invertible} & \multirow{-2}{5em}{Num of Parameters} & \multirow{-2}{3em}{Infer Time} & PSNR & SSIM & PSNR & SSIM \\
\hline
DnCNN~\cite{DnCNN} & \checkmark &  & & & 0.56 & -- & 32.43 & 0.7900 & 23.66 & 0.583 \\ 
\rowcolor{Gray!30}
EPLL~\cite{EPLL} & &  & & & -- & -- & 33.51 & 0.8244 & 27.11 & 0.870 \\ 
% EPLL & Non-Blind & 33.51 & 0.8244 & 27.11 & 0.870 \\ \hline
% \rowcolor{Gray!30}
TNRD~\cite{TNRD}  &  &  & & & -- & -- &  33.65 & 0.8306 & 24.73 & 0.643 \\ 
%NCSR & Non-Blind & 34.05 & 0.8351 \\ \hline
%MLP &  & 34.23 & 0.8331 & 24.71 & 0.641 \\ \hline
\rowcolor{Gray!30}
FFDNet~\cite{zhang2018ffdnet}  &  & &   & & 0.48 & -- & 34.40 & 0.8474 & -- & -- \\ 
BM3D~\cite{BM3D}  &  &  &  & & -- & -- & 34.51 & 0.8507 & 25.65 & 0.685 \\ 
%FoE & Non-Blind & 34.62 & 0.8845 \\ \hline
%WNNM & Non-Blind & 34.67 & 0.8646 & 25.78 & 0.809 \\ \hline
%DenoiseNet & \textcolor{red}{Blind} & 35.08 & 0.868 \\ \hline
\rowcolor{Gray!30}
NI~\cite{NI}  & \checkmark &  & &  & -- & -- &  35.11 & 0.8778 & -- & -- \\ 
NC~\cite{NC}  & \checkmark &  &   & & -- & -- & 35.43 & 0.8841 & -- & -- \\ 
%GCBD & \checkmark & 35.58 & 0.9217 \\ \hline
\rowcolor{Gray!30}
KSVD~\cite{KSVD}  &   & &  & & -- & -- & 36.49 & 0.8978 & 26.88 & 0.842 \\ 
%KSVD-G & -- & -- & 27.19 & 0.771 \\ \hline
%KSVD-DCT & -- & -- & 27.51 & 0.780 \\ \hline
%MCWNNM & & 37.38 & 0.9294 \\ \hline
%FFDNet+ & Non-Blind & 37.61 & 0.9415 \\ \hline
%DnCNN+ & \checkmark & 37.90 & 0.943 \\ \hline
TWSC~\cite{TWSC}  &   & & & & -- & -- & 37.96 & 0.9416 & -- & -- \\ 
\rowcolor{Gray!30}
CBDNet~\cite{Guo2019Cbdnet}  & \checkmark  &  & \checkmark & & 4.34 & 80.76 & 38.06 & 0.9421 & 33.28 & 0.868 \\ 
%PD & \checkmark & 38.40 & 0.9452 \\ \hline
%NLH & \checkmark & 38.81 & 0.952 \\ \hline
%WDnCNN+ & \checkmark & 38.87  & 0.9501  \\ \hline
%ATDNet & \checkmark & 39.19 & 0.9526  \\ \hline
IERD~\cite{CIMM}  &  & & & & -- & -- & 39.20 & 0.9524 & -- & -- \\ 
% MLDN~\cite{}  & \checkmark &  & & & -- &  39.23  & 0.9516 & -- & -- \\ 
\rowcolor{Gray!30}
RIDNet~\cite{RIDNet}  & \checkmark & & & & 1.49 & 196.52 & 39.26 & 0.9528 & -- & -- \\ 
% VDN & \checkmark & 39.38 & 0.9518 \\ \hline
PRIDNet~\cite{PRID_net}  & \checkmark & \checkmark& & & -- & -- & 39.42 & 0.9528 & -- & -- \\ 
\rowcolor{Gray!30}
DRDN~\cite{DRDN}  & \checkmark & & & & -- & -- & 39.43 & 0.9531 & -- & -- \\ 
% DeepProxies BM3D  & & & &--&--& 34.34 & 0.911 \\ \hline
GradNet~\cite{GradNet}  & \checkmark &  & & & 1.60 & -- & 39.44 & 0.9543 & 38.34 & 0.946\\
% ATDNet & \checkmark & 39.19 & 0.9526  \\ \hline
\rowcolor{Gray!30}
% BoostNet~\cite{}  & \checkmark &  & & & $3.76 \times 10^{4}$ &  -- & -- & 38.57 & 0.950 \\
% \rowcolor{Gray!30}
AINDNet~\cite{AINDNet}  & \checkmark & & & & 13.76 & -- & 39.53 & 0.9561 & 39.08 & 0.953 \\ 
DPDN~\cite{DPDN} &\checkmark & & &  & -- & -- & 39.83 & 0.9537 & -- & -- \\ 
% \rowcolor{Gray!30}
\rowcolor{Gray!30}
MIRNet~\cite{Zamir2020MIRNet}  & \checkmark & \checkmark & & & 31.78 & 1569.88 & \textbf{\textcolor{red}{39.88}} & \textbf{\textcolor{red}{0.9563}} & -- & -- \\
VDN~\cite{VDN}  & \checkmark & & & & 7.81 & 99.00 & 39.38 & 0.9518 & 39.26 & 0.955 \\ 
% FAN  & & & & -- & -- & 39.33 & 0.956 \\ \hline
\rowcolor{Gray!30}
DANet~\cite{DANet}  & \checkmark & & \checkmark & & 63.01 & 65.62 & 39.58 & 0.9545 & 39.25 & 0.955 \\ \hline

Ours  & \checkmark & \checkmark & \checkmark & \checkmark & \textcolor{red}{\textbf{2.64}} & \textcolor{red}{\textbf{47.80}} & 39.57 & 0.9522 & \textbf{\textcolor{red}{39.28}} & \textbf{\textcolor{red}{0.955}} \\ \hline
\end{tabular}
}
% 4.35 \times 10^{6}
% \end{centering}
% \vspace{-2mm}
% The mean PSNR and SSIM are reported.%denoising results of algorithms evaluated on the DnD sRGB and SIDD sRGB datasets}
\label{Tab:DnD_SIDD}
% \vspace*{-4mm}
\end{table*}

\subsection{Ablation Study}
\begin{figure}[t]
\centering
% Caption
\begin{subfigure}{.23\textwidth}
  \centering
  \includegraphics[width=\textwidth]{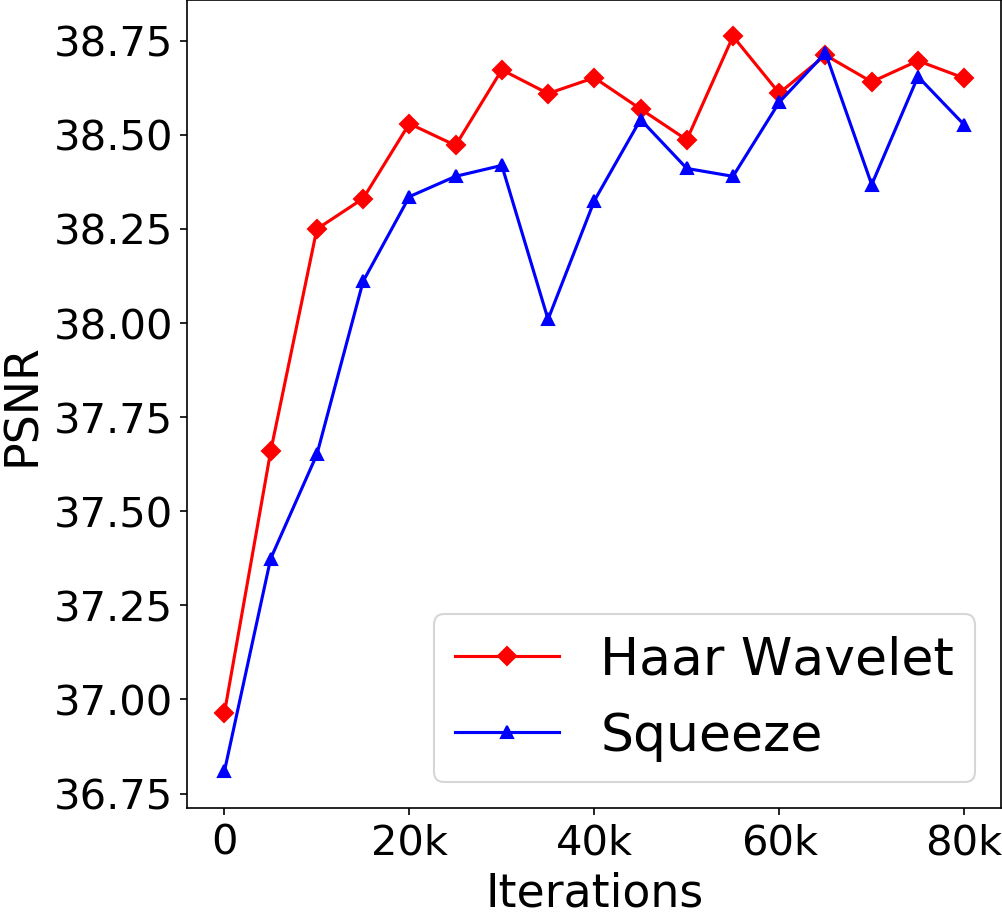} 
  \caption{Transformation}
  \label{subfig:transform}
\end{subfigure}
\begin{subfigure}{.23\textwidth}
  \centering
  \includegraphics[width=\textwidth]{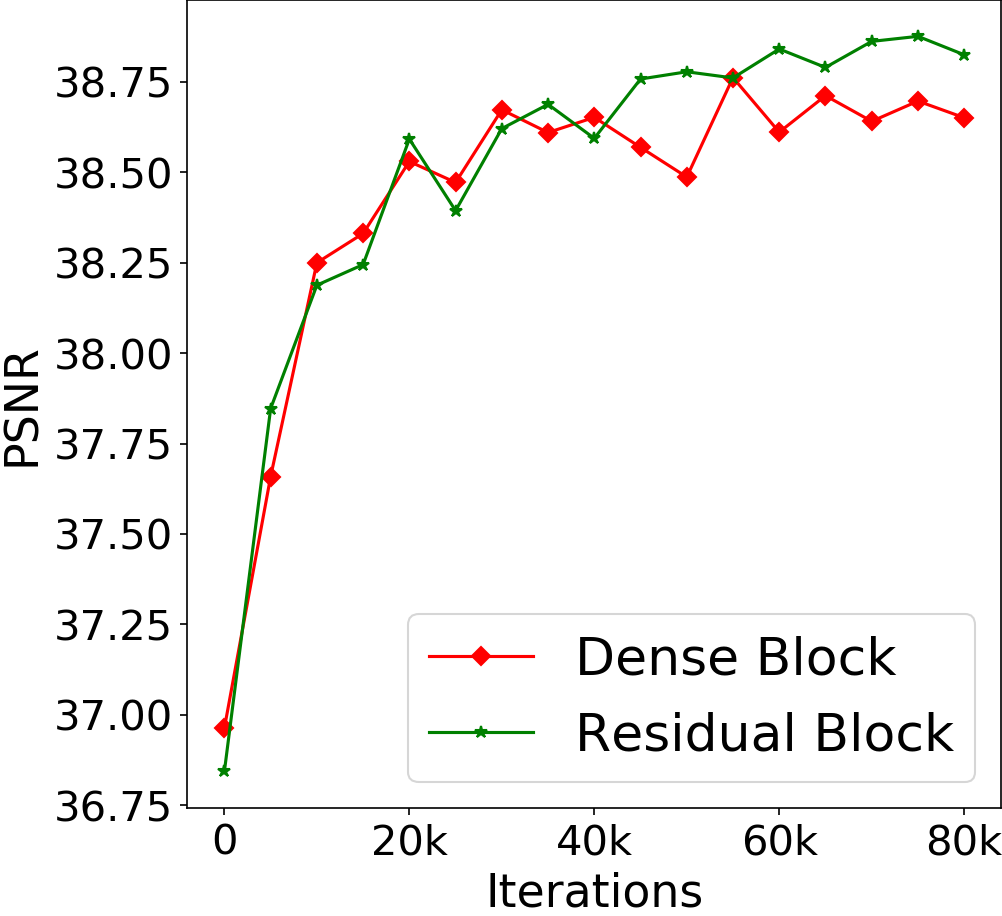} 
  \caption{$\operatorname{\phi}_i$}
  \label{subfig:block}
\end{subfigure}
% \vspace{-2mm}
\caption{Comparisons on training with different network components. (a) illustrates the difference between using Haar wavelet and squeeze as the transformation layer. (b) shows training curves employing residual block and dense block as $\operatorname{\phi}_2$, $\operatorname{\phi}_3$ and $\operatorname{\phi}_4$ separately.}
% \vspace{-3mm}
\end{figure}

\textbf{Squeeze vs Wavelet Transform.}
We compare the difference of employing the squeeze~\cite{kingma2018glow} and Haar wavelet transform in down-scale blocks. We report PSNR on the validation set of the same iteration during training in \autoref{subfig:transform}. The remaining network components and the parameter settings are the same. We observe that using Haar wavelet converges faster and is more stable than the squeeze operation.

\begin{table}[t]
\centering
\caption{Comparisons on the denoising accuracy of different numbers of down-scale blocks and invertible blocks.} 
% \vspace{-2mm}
\begin{tabular}{cccc}
\hline
\hline
Scale & DownScale Blocks & Invertible Blocks & PSNR \\ 
\hline
X2 & 1 & 8 & 38.00   \\ 
X4 & 2 & 8 & \textbf{\textcolor{red}{38.40}}   \\ 
X8 & 3 & 8 & 37.69   \\ \hline
X2 & 1 & 16 & 38.30   \\ 
X4 & 2 & 16 & \textbf{\textcolor{red}{38.61}}   \\ 
X8 & 3 & 16 & 38.15   \\ \hline
\hline
\end{tabular}
\label{Tab:Scale}
% \vspace{-4mm}
\end{table}
\textbf{Residual Block vs. Dense Block.}
Next, we provide comparison between different blocks ($\operatorname{\phi_i}$) in our network in \autoref{subfig:block}. The architecture with residual block achieves higher denoising accuracy than dense block used by~\cite{xiao2020invertible}. Moreover, the network with residual block has far fewer parameters (2.6M) than that using the dense block (4.3M).

\begin{table*}[t]
\centering
\caption{Comparisons on different combinations of losses on the SIDD validation set.}
% \vspace{-3mm}
\resizebox{\textwidth}{!}{
\begin{tabular}{c|cc|cc|cc|cc||cc|cc|cc}
\hline
\hline
\rowcolor{Gray}
Losses & Forw & Back & Forw & Back & Forw & Back & Forw & Back & Forw & Back & Forw & Back & Forw & Back \\ 

\hline
L1 & \checkmark & \checkmark & \checkmark & &  & \checkmark & &  & & \checkmark & & \checkmark & & \checkmark    \\ 
\rowcolor{Gray!30}
L2 &  & & &\checkmark & \checkmark & & \checkmark & \checkmark  & \checkmark &  & \checkmark &  & \checkmark &  \\ 
Gradient Loss &  & & & &  & & & & &\checkmark & & &  & \checkmark \\
\rowcolor{Gray!30}
SSIM Loss & & & & & & & & & & & & \checkmark & & \checkmark\\ \hline
PSNR & \multicolumn{2}{c|}{38.67} & \multicolumn{2}{c|}{38.29} & \multicolumn{2}{c|}{\textbf{\textcolor{red}{38.88}}} & \multicolumn{2}{c||}{38.73} & \multicolumn{2}{c|}{38.74} & \multicolumn{2}{c|}{38.80} & \multicolumn{2}{c}{38.78} \\ \hline
\hline
\end{tabular}
}
\label{Tab:Loss}
% \vspace{-2mm}
\end{table*}

\textbf{Number of Down-Scale and Invertible Blocks. }
% We study the effect of the number of DownScale blocks as well as the number of Invertible Blocks in this part. 
Now, we study the denoising performance of \ourmodel with different numbers of down-scale and invertible blocks. We report the PSNR results on the validation set from the same iteration. We discover that increasing the number of invertible blocks boosts denoising accuracy consistently, regardless of the number of down-scale blocks. Moreover, when fixing the number of invertible blocks, we observe that using two down-scale blocks leads to the best denoising effect.

\textbf{Investigation of Loss Functions.}
\ourmodel requires applying loss functions during training for both the low and high-resolution images to minimize the difference between denoised and clean images. As described in \autoref{subsec:network}, both $\ell_1$- and $\ell_2$- norms are applicable. We investigate InvDN's performance via different combinations of loss functions for producing low and high-resolution images. We compare the PSNR results on the SIDD validation set at the same iteration and report the results in \autoref{Tab:Loss}. We observe that employing $\ell_2$ for low-resolution images and $\ell_1$ for high-resolution images achieves the highest performance. Apart from $\ell_1$ and $\ell_2$, we also explore other loss function candidates to further improve the denoising effects. We apply the gradient loss~\cite{GradNet} and the SSIM loss~\cite{7797130} to minimize the difference between two images from the perspectives of image derivative and structures. However, neither the gradient nor the SSIM loss improves the denoising performance.

\begin{figure}[t]
\captionsetup[subfigure]{justification=centering}
 \centering
\begin{subfigure}{.15\textwidth}
  \centering
  % include second image
  \includegraphics[width=.99\linewidth]{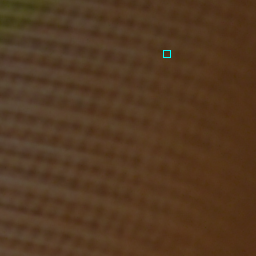}  
  \caption{Ground-\\Truth}
  \label{subfig:noise_gt}
\end{subfigure}
\begin{subfigure}{.15\textwidth}
  \centering
  % include second image
  \includegraphics[width=.99\linewidth]{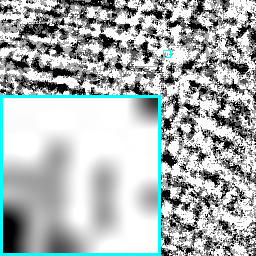} 
  \caption{Before MC Self-Ensemble}
  \label{subfig:noise_z_1}
\end{subfigure}
\begin{subfigure}{.15\textwidth}
  \centering
  % include second image
  \includegraphics[width=.99\linewidth]{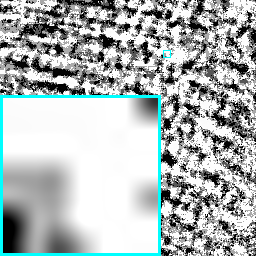}  
  \caption{After MC Self-Ensemble}
  \label{subfig:noise_z_16}
\end{subfigure}
% \vspace{-1mm}
\caption{Visualization of denoising performance before and after MC self-ensemble. (b) and (c) reflect the difference between the denoised image and the ground-truth. Darker means more difference, indicating poorer denoising. }
\label{fig:mc_self_ensem}
% \vspace{-1mm}
\end{figure}

\begin{table}[t]
    \centering
    \caption{AKLD of generated noisy images by different models. Lower is better. The lowest AKLD is in red. }
    % \vspace{-1mm}
    \resizebox{0.48\textwidth}{!}{
    \begin{tabular}{c c c c c c}
    \toprule
         CBDNet & ULRD & GRDN & Noise Flow & DANet & \ourmodel \\
         \cite{Guo2019Cbdnet} & \cite{ULRD} & \cite{GRDN} & \cite{9008378} & \cite{DANet} & (Ours) \\ 
         \midrule
         0.728 & 0.545 & 0.443 & 0.312 & 0.212 & \textbf{\textcolor{red}{0.059}} \\
    \bottomrule
    \end{tabular}
    }
    \label{tab:akld}
    % \vspace{-4mm}
\end{table}

% \subsection{Monte Carlo Self Ensemble}
% \label{subsec:mc_self_ensemble}
% To further improve \ourmodels's denoising effect without using extra data and training, we introduce a novel self-ensembling method. As we have described in \autoref{subsec:concep_of_design}, when reverting a low-resolution image to the original input's size, we randomly sample a latent variable $\zhf$ from Gaussian distribution $\ngaussian$. However, the reverted image may slightly deviate from the ground-truth clean image but follows a distribution whose mean is the ground-truth. Therefore, to reduce these deviations, we utilize the Monte Carlo method by sampling the latent variable $\zhf$ multiple times, resulting in a set of latent variables $\{ \zhf^i \}_{i=1}^{N}$, where for each $\zhf^i$, a corresponding denoised image $\hat{x}_i$ is obtained. The final output is the average of $\{ \hat{x}_{i} \}_{i=1}^{N}$. By the law of large numbers~\cite{wu2018strong}, the averaged image is closer to the ground-truth than any individual $\hat{x}_i$. Because the Monte Carlo method does not require additional supervision, we name it as Monte Carlo (MC) self-ensemble. 

% \textbf{Monte Carlo Self Ensemble. }
\subsection{Monte Carlo Self Ensemble}
To further improve \ourmodels's denoising effect without using extra data and training, we introduce Monte Carlo (MC) self-ensemble. 
% The denoising performance of \ourmodel can thereby be further improved, which does not use extra data and training. 
We sample the latent variable $\zhf$ multiple times, resulting in a set of latent variables $\{ \zhf^i \}_{i=1}^{N}$, where for each $\zhf^i$, a corresponding denoised image $\hat{x}_i$ is obtained. The final output is the average of $\{ \hat{x}_{i} \}_{i=1}^{N}$. 
By the law of large numbers~\cite{wu2018strong}, the averaged image is closer to the ground-truth than any individual $\hat{x}_i$.
In \autoref{fig:mc_self_ensem}, we visualize the denoising results before and after using MC self-ensembling by setting the MC size to 16. The contrast between the residual images before and after MC self-ensemble indicates that it further reduces the noise in the denoised image. Quantitatively, 83.35\% images witness a performance boost on the SIDD validation set.
\label{subsec: study_of_z}

\subsection{Analysis of Distributions of $\z$}
\begin{figure}[t]
\captionsetup[subfigure]{justification=centering}
\centering
\begin{subfigure}{.15\textwidth}
  \centering
  % include first image
  \includegraphics[width=.99\linewidth]{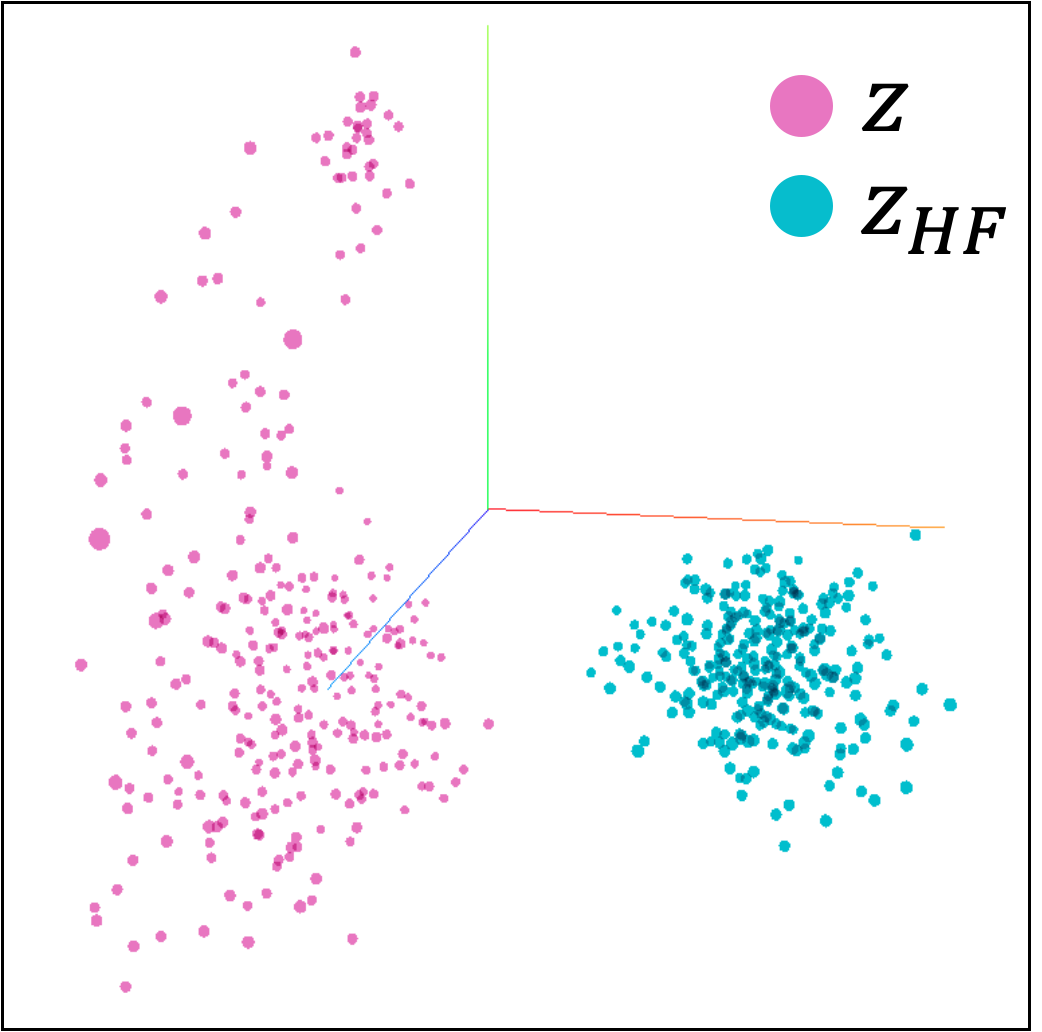}  
  \caption{Distributions}
  \label{subfig:z_two_dists}
\end{subfigure}
\begin{subfigure}{.15\textwidth}
  \centering
  % include first image
  \includegraphics[width=.99\linewidth]{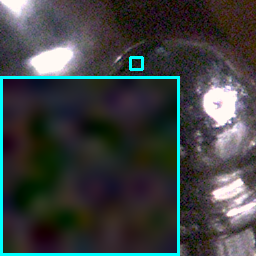}  
  \caption{Original}
  \label{subfig:noisy_orig}
\end{subfigure}
\begin{subfigure}{.15\textwidth}
  \centering
  % include first image
  \includegraphics[width=.99\linewidth]{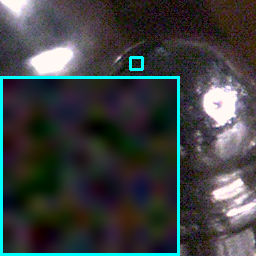} 
  \caption{Generated}
  \label{subfig:noisy_deviation}
\end{subfigure}
% \vspace{-1mm}
\caption{(a) PCA visualization results of the two distributions corresponding to the noisy latent representations $\z$ (pink) and the clean samples $\zhf$ (blue). (b) Original noisy image from the SIDD validation set. (c) The generated noisy image by \ourmodel corresponding to (b). }
% \vspace{-2mm}
\end{figure}

For \ourmodels, there are two types of latent variables: one corresponds to the original high-frequency signal $\z$ containing noise, and the other sampled $\zhf$ from $\ngaussian$ representing clean details. To analyze, we obtain 500 pairs of $\z$ and $\zhf$ from the SIDD validation set. We vectorize each latent variable and plot them in a 3D space with PCA~\cite{PCA}. As \autoref{subfig:z_two_dists} shows, $\z$ and $\zhf$ follow two different distributions.
% , indicating the validity of \ourmodel in removing noise. 

For the sake of fair comparison, we only train \ourmodel on the benchmark SIDD training set. However, \ourmodel also supports generating more noisy images for data augmentation. To generate augmented data, we first conduct sampling $\z$. We introduce a tiny disturbance to $\z$ as $\z' = \z + \epsilon \cdot \mathbf{v}$, where $\mathbf{v} \sim \ngaussian$ and $\epsilon$ is set as $2 \times 10^{-4}$. We expect the reverted image of $\z'$ to have the same background clean image, yet with a different noise. We visualize the reverted images of $\z$ and $\z'$ in \autoref{subfig:noisy_orig}~and~\ref{subfig:noisy_deviation}, exhibiting visually different noise corresponding to $\z$ and $\z'$. 

To quantitatively evaluate the quality of our generated noisy images, we follow the average KL divergence (AKLD) metric introduced by DANet~\cite{DANet} to measure the similarity between the original and the generated noise, as shown in \autoref{tab:akld}. It should be noted here that  Noise Flow~\cite{9008378} requires raw-RGB images, ISO, and CAM information to generate noisy images; in other words, it needs the training images as inputs.
The AKLD results in \autoref{tab:akld} demonstrate that our generated noisy images are closer to the original noisy images by a large margin.

\section{Conclusion}
This paper is the first to study real image denoising with invertible networks. In previous invertible models, the input and the reversed output follow the same distribution.  However, for image denoising, the input is noisy, and the restored outcome is clean, following two different distributions. To address this issue, 
our proposed \ourmodel transforms the noisy input into a low-resolution clean image as well as a latent representation containing noise. 
As a result, \ourmodel can both remove and generate noise. For noise removal, we replace the noisy representation with a new one sampled from a prior distribution to restore clean images; 
for noise generation, we alter the noisy latent vector to reconstruct new noisy images. 
Extensive experiments on three real-noise datasets demonstrate the effectiveness of our proposed model in both removing and generating noise. 
% Extensive experiments on three real-noise datasets demonstrate the effectiveness of our proposed model.

\vspace{.5em}

\small{%feel free to add more
\noindent \textbf{Acknowledgments}: 
This work was partly supported by Institute of Information \& communications Technology Planning \& Evaluation (IITP) grant funded by the Korea government(MSIT) (No.2019-0-01906, Artificial Intelligence Graduate School Program(POSTECH)).
}

% \newpage
% \onecolumn
% \appendix
% \input{CVPR2021/supp/fig_sidd}

\end{document}